\documentclass{elsart}
\usepackage{graphicx}
\usepackage{dcolumn}
\usepackage{bm}
\usepackage{amsmath}
\usepackage{amsfonts}
\usepackage{amssymb}
\providecommand{\U}[1]{\protect\rule{.1in}{.1in}}

\begin{document}
\begin{frontmatter}
\title{Anomalous diffusion in the resonant quantum kicked rotor}
\author{A. Romanelli, }
\author{Guzm\'an Hern\'andez}
\address{Instituto de F\'{\i}sica, Facultad de Ingenier\'{\i}a\\
Universidad de la Rep\'ublica\\ C.C. 30, C.P. 11000, Montevideo,
Uruguay\\
e-mail: alejo@fing.edu.uy, guzmanhc@fing.edu.uy}
\date{\today}
\begin{abstract}
We study the resonances of the quantum kicked rotor subjected to an
excitation that follows a deterministic time-dependent prescription.
For the primary resonances we find an analytical relation between
the long-time behavior of the standard deviation and the external
kick strength. For the secondary resonances we obtain essentially
the same result numerically. Selecting the time sequence of the kick
allows to obtain a variety of asymptotic wave-function spreadings:
super-ballistic, ballistic, sub-ballistic, diffusive, sub-diffusive
and localized.
\end{abstract}
\begin{keyword}
Quantum resonance; Kicked rotor; Anomalous diffusion
\end{keyword}
\end{frontmatter}

\section{Introduction}

The kicked rotor model is a paradigm of periodically driven systems
with classical chaotic behavior and its quantum version, the quantum
kicked rotor (QKR), can be considered a cornerstone in the study of
chaos at the quantum level \cite{CCI79}. This topic is a matter of
permanent attention, both theoretical and experimental, for example
it has been recently reported \cite{Chaudhury} the experimental
realization of the quantum kicked top (a variant of the QKR.) to
study signatures of quantum chaos.

The behavior of the QKR has two characteristic modalities: dynamical
localization and ballistic spreading of the variance in resonance
\cite{Izrailev}. These behaviors are quite different and have no
classical analog. They depend on the relationship between the
characteristic time of the free rotor and the period associated to
the kick. When the dimensionless period of the kick, $T$, is an
irrational multiple of $2\pi$ the average energy of the system grows
in a diffusive manner for a short time and afterwards dynamical
localization appears. In this case, the angular momentum
distribution has a maximum at its mean value and decays
exponentially. For $T$ rational multiple of $2\pi$ the behavior of
the system is resonant with ballistic spreading, and the angular
momentum distribution evolves from the initial one in such away that
its standard deviation has the time dependence $\sigma(t)\sim t$.
The two types of values of $T$ determine the spectral properties of
the Hamiltonian, for irrational multiples the energy spectrum is
purely discrete and for rational multiples it contains a continuous
part. Both quantum resonance and dynamical localization can be seen
as interference phenomena, the first being a constructive
interference effect and the second a destructive one.

Different variants of the QKR have been studied, but with the aim of
this work in mind, it is interesting to highlight the one developed
by \cite{Casati,victor} where the QKR is subjected to two unitary
operators with two different values of the strength parameter $K$.
The dynamics of the QKR kicked according to a Fibonacci sequence is
studied in Ref. \cite{Casati}. This is done outside the resonant
regime, where a sub-diffusive behavior is found for small kicking
strengths. In Ref. \cite{victor} we have also studied the QKR
excited with the same Fibonacci prescription but in the resonant
regime. It was shown that the standard deviation of the primary
resonances maintains the well-known ballistic behavior but the
secondary resonances have sub-ballistic behavior, characterized by
the time dependence of its standard deviation $\sigma(t)\sim
t^{\gamma}$, with $1/2<\gamma<1$. This change of behavior is similar
to that obtained for the quantum walk (QW) subjected to an aperiodic
Fibonacci excitation \cite{Ribeiro} where the ballistic behavior is
lost and the sub-ballistic behavior appears. We have also
investigated the QKR in resonance subjected to decoherence: a) with
a L\'evy waiting-time distribution \cite{alejo1,alejo2} and b) with
an unitary noise defined through Kraus operators \cite{alejo5}. In
both cases we find that the secondary resonances have a
sub-ballistic behavior, while the principal resonances are not
affected. On the other hand, in Refs. \cite{Schomerus1,Schomerus2}
the QKR subjected to noises with a L\'evy distribution out of
resonance was investigated, showing that the decoherence leads to a
sub-diffusive regime ($\gamma<1/2$) for a short time before
localization appears.

The quantum resonances and the dynamical localization have been
experimentally observed in Refs. \cite{Moore,Bharucha,Raizen,Kanem}.
The QKR was experimentally realized through a dilute sample of
ultra-cold atoms exposed to a one-dimensional spatially periodic
optical potential that is pulsed on periodically in time (to
approximate a series of delta function kicks). However, the
ballistic growth of the secondary resonances was not confirmed. It
was found that the momentum distribution profiles saturated to a
final distribution but different from the characteristic exponential
profiles of the dynamical localization. In Ref. \cite{Bharucha} the
usual QKR is modified, by adding other elements such as a
quasi-momentum, in order to explain the absence of the ballistic
growth. However later experiments \cite{Bharucha} has confirmed this
growth.

In a recent paper \cite{monva} we have studied the QW model using a
time-dependent unitary coin operator. We found different types of
anomalous diffusion depending on the time prescription for the coin
operator. In the same line of thought, in this paper we develop a
simple QKR model with a generalized time strength parameter $K(t)$
that allows an analytical treatment of the primary resonances. We
study the connection between the time evolution of $K(t)$ and the
type of spreading of the system. We analytically show how to select
the time dependence of $K(t)$ to obtain a predetermined power-law
distribution of the standard deviation. We propose this toy model to
reproduce the experimental behavior reported in Refs.
\cite{Bharucha,Kanem}.

The paper is organized as follows: In the next section we present a
brief revision of the QKR equations, in the third section we obtain
analytically the standard deviation for the primary resonances, in
section $IV$ we choose a time dependence for $K(t)$, in section $V$
numerical results are presented for the secondary resonances, and in
the last section we draw the conclusions.

\section{Quantum kicked rotor}

The QKR is one of the simplest and best investigated quantum models
whose classical counterpart displays chaos. In this paper we use a
modified version of the usual QKR where a deterministic time
dependence for the strength parameter $K(t)$ is introduced. The
Hamiltonian has then the following shape
\begin{equation}
H(t)=\frac{P^{2}}{2I}+K(t)\cos \theta \sum_{n=1}^{\infty }\delta
(t-nT), \label{qkrham}
\end{equation}%
where the external kicks occur at times $t=nT$ with integer $n$, $T$
being the kick period (the index $n$ will be equivalent to time in
units of $T$, given the delta function appearing in
Eq.(\ref{qkrham})), $I$ is the moment of inertia of the rotor, $P$
the angular momentum operator and $\theta$ the angular position
operator. The last two operators satisfy the canonical commutation
rule
\begin{equation}
[P,\theta]=-i\hbar. \label{commuta}
\end{equation}
In the angular momentum representation, $P|\mathit{l} \rangle
=\mathit{l} \hbar |\mathit{l} \rangle $, the wave-vector is
\begin{equation}
|\Psi (n)\rangle =\sum_{\mathit{l} =-\infty }^{\infty }a_{\mathit{l} }(n)|%
\mathit{l} \rangle  \label{psi}
\end{equation}
and the average energy is
\begin{equation}
E(n)=\left\langle \Psi \right\vert H\left\vert \Psi \right\rangle
=\varepsilon \sum_{\mathit{l} =-\infty }^{\infty }\mathit{l} ^{2}\left\vert
a_{\mathit{l} }(n)\right\vert ^{2},  \label{ene}
\end{equation}
where $\varepsilon =\hbar ^{2}/2I$. Using the Schr\"{o}dinger
equation the quantum map is readily obtained from the Hamiltonian
Eq.(\ref{qkrham})
\begin{equation}
a_{\mathit{l} }(n+1)=\sum_{j=-\infty }^{\infty }U_{\mathit{l}
j}a_{j}(n), \label{mapa}
\end{equation}%
where the matrix element of the time step evolution operator
$U(\kappa)$ is
\begin{equation}
U_{\mathit{l} j}=i^{-(j-\mathit{l} )}e^{-ij^{2}\varepsilon T/\hbar }\,J_{j-%
\mathit{l} }(\kappa(n)),  \label{evolu}
\end{equation}%
$J_{\mathit{l}}$ being the $\mathit{l}$th order cylindrical Bessel
function and its argument the dimensionless kick strength
$\kappa(n)\equiv K(n)/\hbar$. With the aim to generate the dynamics
of the system we will consider different time dependencies of the
strength parameter and combine the corresponding time-step operators
$U\left(\kappa(n)\right)$ in a large sequence. The resonance
condition does not depend on $\kappa$ and takes place when the
frequency of the driving force is commensurable with the frequencies
of the free rotor. Inspection of Eq.(\ref{evolu}) shows that the
resonant values of the scale parameter $\tau\equiv \varepsilon
T/2\hbar$ are the set of the rational multiples of $2\pi$,
$\tau=2\pi$ $p/q$. In what follows we assume, that the resonance
condition is satisfied, therefore the evolution operator depends on
$\kappa$, $p$ and $q$. We call a resonance primary when $p/q$ is an
integer and secondary when it is not.

\section{Primary resonances}

We shall first consider the primary resonances. Using the recursion
relation satisfied by the Bessel functions Eq.(\ref{mapa}) is
solved. Its general solution is written as
\begin{equation}
a_{\mathit{l}}(n)=\sum\limits_{j=-\infty }^{\infty }\left( -i\right) ^{%
\mathit{l}-j}a_{j}(0)\text{ }\,J_{\mathit{l}-j}(n^{\ast }),  \label{solua}
\end{equation}%
where $a_{j}(0)$ are the initial amplitudes and
\begin{equation}
n^{\ast }=\sum\limits_{m=1}^{n}\kappa (m).  \label{time}
\end{equation}%
From Eq.(\ref{solua}) it is clear that the propagation speed of the
probability amplitudes is given by $\Delta n^{\ast }/\Delta n=\kappa
(n)$. The probability distribution at time $n$ is given by
\begin{equation}
P_{\mathit{l}}(n)=|a_{\mathit{l}}(n)|^{2},  \label{prob}
\end{equation}%
that can be expressed as
\begin{equation}
P_{\mathit{l}}(n)=\sum_{j,k=-\infty }^{\infty }\left(-i\right)
^{k-j}
a_{k}(0)a_{j}^{\ast }(0) J_{\mathit{l}-k}(n^{\ast })J_{\mathit{l}%
-j}(n^{\ast }).  \label{prob1}
\end{equation}%
Using the  properties of the Bessel functions all the moments of
$P_{\mathit{l}}(n)$ can be calculated analytically from
Eq.(\ref{prob1}). If symmetrical initial conditions are taken, the
odd moments vanish due to the symmetrical dependence on the initial
conditions. We calculate the first and second moments, obtaining:
\begin{equation}
M_{1}(n)=-n^{\ast }\sum_{j=-\infty }^{\infty }\Im \left[ a_{j}(0)a_{j-1}^{%
\ast }(0)\right] +M_{1}(0), \label{mome0}
\end{equation}
\begin{eqnarray}
M_{2}(n) &=&\frac{(n^{\ast })^{2}}{2}\left( 1-\sum_{j=-\infty }^{\infty }\Re %
\left[ a_{j}(0)a_{j+2}^{\ast }(0)\right] \right)  \notag \\
&&+n^{\ast }\sum_{j=-\infty }^{\infty }\left( 2j+1\right) \Im \left[
a_{j}(0)a_{j+1}^{\ast }(0)\right]  \notag \\
&&+M_{2}(0),  \label{mome}
\end{eqnarray}%
where $\Re \left[ x\right]$ and $\Im \left[ x\right]$ are the real
and imaginary part of $x$ respectively. The standard deviation
$\sigma =\sqrt{M_{2}-M_{1}^{2}}$ has the following long-time limit
\begin{equation}
{\sigma }(n)\rightarrow C n^{\ast }=C \sum\limits_{m=1}^{n}\kappa
(m), \label{sigma}
\end{equation}%
where $C$ is a constant that only depends on the initial conditions.
\section{Choosing $K(t)$}
Among the alternatives to choose the time dependence of $\kappa(n)$
we study the case:
\begin{equation}
\kappa (n)=\kappa_{0}\left( \frac{1}{n}\right) ^{\alpha },
\label{cos}
\end{equation}
where $\alpha$ is a real parameter and $\kappa_{0}$ is a constant.
Substituting Eq.(\ref{cos}) in Eq.(\ref{sigma}) and omitting a
multiplicative constant the following asymptotic results are
obtained
\begin{equation}
{\sigma }(n)\rightarrow \left\{
\begin{array}{c}
\ln n,\text{ if }\alpha =1, \\
\\
n^{\gamma },\text{ if }\alpha \neq 1,%
\end{array}%
\right.   \label{sigmaf}
\end{equation}%
where
\begin{equation}
\gamma \equiv 1-\alpha. \label{gamma}
\end{equation}
The asymptotic spreading of the wave function shows seven different
behaviors that can be characterized by the exponent $\gamma$:

(a) $\gamma >1$, super-ballistic.

(b) $\gamma =1$, ballistic.

(c) $1>\gamma >1/2$, sub-ballistic.

(d) $\gamma =1/2$, diffusive.

(e) $1/2>\gamma >0$, sub-diffusive.

(f) $\gamma =0$, logarithmic.

(g) $\gamma <0$, Bessel-localized.

The result for $\gamma =1$ is expected because the system reduces to
the usual QKR in resonance. In the other cases the system shows
unexpected behaviors. For $\gamma <0$ the system does not spread
when $t\rightarrow \infty$, it becomes localized in angular
momentum. We have called this behavior `Bessel-localized' because
the distribution decay is slower than the case of exponential
localization, as can be seen from Eq.(\ref{prob1}). Note that this
equation with initial conditions $a_k(0)=\delta_{k m}$ is reduced
to:
\begin{equation}
P_{\mathit{l}}(n)= \left[{J_{\mathit{l}-m}}(n^{\ast })\right]^2.
\label{prob2}
\end{equation}
In the other extreme for $\gamma>1$ the spreading of the
wave-function is super-ballistic. In this case not only the external
frequency matches the natural frequency of the free rotator
(resonance condition) but the intensity of the external strength
grows with time as well. We could think that this condition produces
a new kind of reinforced resonance.
\begin{figure}[th]
\begin{center}
\includegraphics[scale=0.38]{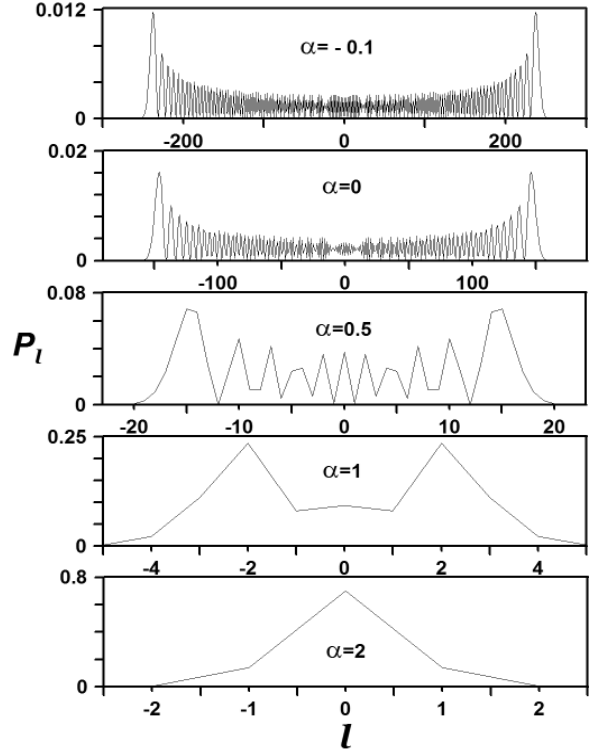}
\end{center}
\caption{The probability distribution $P_{\mathit{l}}$, for a
primary resonance as a function of the dimensionless momentum
$\mathit{l}$ at time $n=300$ with the initial condition
$a_\mathit{l}(0)=\delta_{\mathit{l} 0}$. The values of the parameter
$\alpha$ are, from top to bottom, $-0.1,\,0,\,0.5,\,1$ and $2$.}
\label{f0}
\end{figure}
In order to visualize the asymptotic behavior of the distribution
$P_{\mathit{l}}$, we use Eq.(\ref{prob2}) to calculate the profiles
of the distribution. These profiles are shown in Fig.~\ref{f0} for
several values of $\alpha$ after a time long enough has elapsed, to
see the spreading of the distribution. We take the initial
conditions $a_{\mathit{l}}(0)=\delta _{\mathit{l}0}$, to produce a
symmetrical evolution. From this figure, it is clear that as
$\alpha$ grows the spreading speed decreases, the distribution
shrinks and the two extreme peaks come closer. When $\alpha =1/2$
the standard deviation spreads out as $\sigma (t)\sim t^{1/2}$, but
the probability distribution $P_{\mathit{l}}$ is not Gaussian. This
confirms that the evolution corresponds to a coherent unitary
process and not to the classical diffusive behavior. In the extreme
case when $\alpha>1$ the distribution is restricted to a small
interval around its initial value and the two extreme peaks melt
into a very narrow peak. It is important to point out that the
distribution profiles presented here are qualitatively similar to
those obtained in Ref. \cite{banuls,monva} for the QW.

To conclude this section, it is important to point out that the
specific form of the kick sequences in Eq.(\ref{cos}) is not
essential to obtain the qualitative behaviors described above, they
are independent of this choice. Note that ${\sigma }(n)$ is
determined by Eq.(\ref{sigma}) and the qualitative behavior of
$\kappa(n)$. If $\kappa(n)\rightarrow 0$, ${\sigma }(n)$ cannot grow
faster than in the ballistic case, therefore it behaves as in the
cases denoted above by (b), (c), (d), (e) or (g), depending on the
rate of decay of $\kappa(n)$. When $\kappa(n) \nrightarrow 0$ but
remains bounded, the behavior is ballistic. Finally when $\kappa(n)$
is not bounded, ${\sigma }(n)$ grows super-ballistically as in case
(a).
\section{Secondary resonances}

Now we inquire about the secondary resonances of the QKR. In the
general case the analytical development is very cumbersome because
the matrix elements of the unitary map in Eq.(\ref{evolu}) include
the time dependent phases $ e^{-ij^{2}\varepsilon T/\hbar }$. These
phases impede the integrability of the problem. It could be possible
to obtain analytical results for some special function $\kappa(t)$
following the work of Ref. \cite{IzraiandShepe}. However, we opt to
study the secondary resonances numerically iterating Eq.(\ref{mapa})
with the time dependent kick proposed in Eq.(\ref{cos}) to calculate
the distribution $ P_{\it l}$ and the standard deviation.
\begin{figure}[th]
\begin{center}
\includegraphics[scale=0.38]{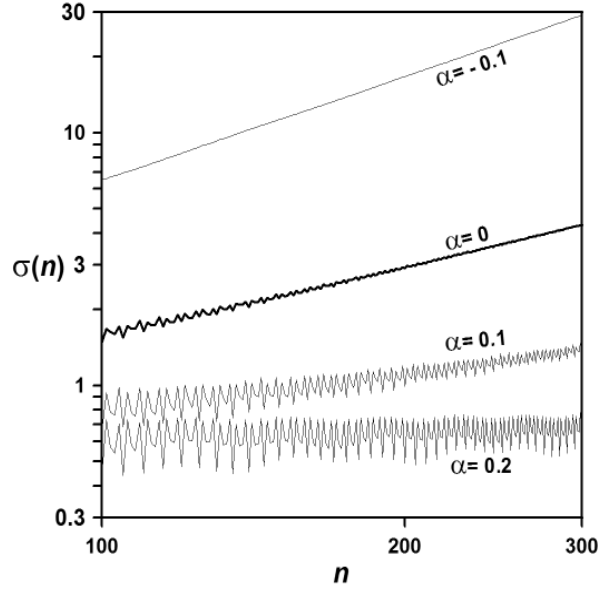}
\end{center}
\caption{The standard deviation ${\protect\sigma }(n)$ as a function
of the dimensionless time $n$ in log-log scales for the secondary
resonance $p/q=2/5$. The initial condition for the amplitudes is
$a_\mathit{l}(0)=\delta_{\mathit{l} 0}$.
For large $n$ the curves satisfy, on the average, a power law ${{n}^{\protect%
\gamma }}$. The values of $\protect\gamma $ are, from top to bottom, (1) $%
\protect\gamma \sim 1.4$ for $\protect\alpha =-0.1$, (2) $\protect\gamma =1$
for $\protect\alpha =0$, (3) $\protect\gamma \sim 0.5$ for $\protect\alpha %
=0.1$, and (4) $\protect\gamma \sim 0$ for $\protect\alpha =0.2.$}
\label{f1}
\end{figure}
In Fig.~\ref{f1} the standard deviation $\sigma$ for the secondary
resonance $p/q=2/5$ is presented for several values of the parameter
$\alpha$. It is seen that $\sigma$ has a power-law growth with an
exponent $\gamma$ that depends on $\alpha$, but now the relation
given by Eq.(\ref{gamma}) does not hold. The values of $\gamma$ were
adjusted for the last hundred values of $n$. The ranges of $\alpha$
for the different behaviors are now the following:

(a) $\alpha<0$, super-ballistic ($\gamma >1$),

(b) $\alpha=0$, ballistic ($\gamma =1$),

(c) $0<\alpha \lesssim 0.1$, sub-ballistic ($1>\gamma \gtrsim 0.5$),

(d) $\alpha \simeq 0.1$, diffusive ($\gamma \simeq 0.5$),

(e) $0.1\lesssim \alpha \lesssim 0.2$, sub-diffusive $0.5\gtrsim
\gamma \gtrsim 0$,

(f) $\alpha \gtrsim 0.2$, localized ($\gamma \lesssim 0$).

If we compare these ranges with the corresponding ranges of $\alpha$
for the primary resonances we find a similar qualitative behavior,
as in the cases (a), (b), (c), (d) and (e). However in the present
calculation, the logarithmic case is not present and it is not easy
to classify the type of localization obtained.

\begin{figure}[th]
\begin{center}
\includegraphics[scale=0.38]{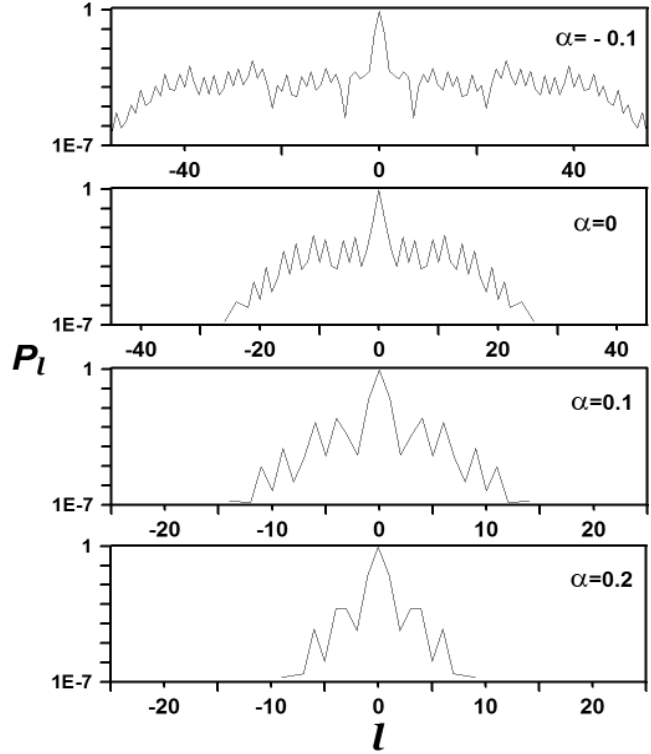}
\end{center}
\caption{The probability distribution $P_{\mathit{l}}$, for the
secondary resonance $p/q=2/5$, as a function of the dimensionless
momentum $\mathit{l}$ at time $n=300$ for the initial condition
$a_\mathit{l}(0)=\delta_{\mathit{l} 0}$. The vertical scale is
logarithmic.} \label{f2}
\end{figure}

In Fig.~\ref{f2} we present the distribution $P_{\mathit{l}}$ as a
function of the angular momentum for four values of $\alpha$. The
distributions show a symmetrical spreading around a central peak.
The spreading shrinks as $\alpha$ grows, going towards the localized
behavior, as in the case of the primary resonances.
\section{Discussion and conclusion}

The resonances of the QKR, as well as the dynamical localization,
were first established numerically \cite{CCI79} and the analytical
treatment of the resonances was developed by Izrailev and
Shepelyansky \cite{IzraiandShepe}. Thirteen years ago the primary
resonances were experimentally observed by Bharucha \emph{et al.}
\cite{Bharucha}, in samples of cold atoms interacting with a
far-detuned standing wave of laser light. More recently Kanem
\emph{et al.} \cite{Kanem} have observed secondary resonances. These
notable series of experiments, together with those in Refs.
\cite{Moore,Raizen,Ammann}, have drawn much attention because they
have allowed to study fundamental aspects of the interaction between
matter and radiation, and may also contribute to the experimental
basis of quantum computation. In Ref. \cite{Bharucha} the usual QKR
model was used to describe the experimental results where dynamical
localization was found out of resonance. However, they did not
observe the ballistic growth in resonance predicted by the model.
Instead, an unexpected decay of the distribution was obtained. The
variety of anomalous behaviors of the resonances presented in this
paper could be useful to interpret these experimental results.

On the other hand, the experimental set up described in the previous
references could be used for the realization of the QKR proposed in
this work. The most important difference with the realized
experiments would reside in that the intensity of the laser that
hits the cloud of atoms should depend on time.

In summary, we have developed a variant of the QKR model that shows
anomalous diffusion in resonances. The spreading of the wave
function has a rich variety of behaviors that go from the localized
to the super-diffusive. This feature opens interesting possibilities
for quantum information processing because it could be used to
control the spreading of the wave function. For the primary
resonances, in the long-time asymptotic regime, we have found an
analytical expression for the dependence of the standard deviation
with the strength parameter $K(t)$. This expression shows explicitly
how to select the unitary evolution to obtain a predetermined
asymptotic behavior for the wave function spreading:
super-ballistic, ballistic, sub-ballistic, diffusive, sub-diffusive,
logarithmic or Bessel-localized. The parameter $\alpha$ determines
the degree of diffusivity of the system in the long-time limit. We
have shown numerically that the standard deviation for the secondary
resonances has the same qualitative behavior as that exhibited by
the primary resonances. The probability distributions are similar in
the case of localization for both types of resonances, but in the
other cases the profiles are very different.

Finally, it is important to mention that there exists a parallelism
between the results obtained in the present work and those obtained
in Ref. \cite{monva} for the QW. This evidences the theoretical
equivalence between the QKR in resonance and the QW
\cite{alejo1,auyuanet,alejo6}.

We acknowledge the support from PEDECIBA, ANII and thank V.
Micenmacher for his comments and stimulating discussions.

\end{document}